\documentclass[journal]{IEEEtran}
\usepackage{bm}
\usepackage{multicol}
\usepackage{mathtools}
\usepackage{amssymb}
\usepackage{amsmath}
\usepackage{subeqnarray}
\usepackage{color}

\usepackage[dvips]{graphicx}
\usepackage{subfigure}
\usepackage{array}
\usepackage{url}
\usepackage{algorithm}
\usepackage{algorithmic}
\usepackage{amsthm}
\theoremstyle{plain}
\usepackage{color}

\theoremstyle{definition}

\theoremstyle{remark}

\usepackage{cite}

\begin{document}
\title{Preserving Reliability to Heterogeneous Ultra-Dense Distributed Networks in Unlicensed Spectrum}
\vspace{1cm}
\author{{Qimei Cui, Yu Gu, Wei Ni, Xuefei Zhang, Xiaofeng Tao, Ping Zhang, and Ren Ping Liu}
\thanks{Qimei Cui, Yu Gu, Xuefei Zhang, Xiaofeng Tao and Ping Zhang are with Beijing University of Posts and Telecommunications; Wei Ni is with Commonwealth Scientific and Industrial Research Organization, Sydney, Australia, NSW 2122; Ren Ping Liu is with University of Technology Sydney, Australia, NSW 2000.}}
\maketitle
\begin{abstract}
This article investigates the prominent dilemma between capacity and reliability in heterogeneous ultra-dense distributed networks, and advocates a new measure of effective capacity to quantify the maximum sustainable data rate of a link while preserving the quality-of-service (QoS) of the link in such networks.
Recent breakthroughs are brought forth in developing the theory of the effective capacity in heterogeneous ultra-dense distributed networks. Potential applications of the effective capacity are demonstrated on the admission control, power control and resource allocation of such networks, with substantial gains revealed over existing technologies.
This new measure is of particular interest to ultra-dense deployment of the emerging fifth-generation (5G) wireless networks in the unlicensed spectrum, leveraging the capacity gain brought by the use of the unlicensed band and the stringent reliability sustained by 5G in future heterogeneous network environments.
\end{abstract}


\section{Introduction}
Future wireless networks are anticipated to be deployed in an ultra-dense fashion to boost network capacity. This is because
the ultra dense deployment is able to compensate for the
limited bandwidth by reusing the bandwidth geographically. The ultra dense deployment could be implemented in a centralized manner, such as Cloud Radio Access Network (C-RAN)~\cite{ni2013new}, or a hybrid fashion with centralized management of traffic routing and distributed interference avoidance control~\cite{Shangjing2015},
or more desirably in a completely distributed manner~\cite{Cognitivevehicular}.
Particularly, the distributed deployment is envisaged as such that inexpensive miniature cellular base stations can be ultra-densely installed in an uncoordinated fashion, adapting to the geographically varying distribution of traffic demand. These base stations are expected to automate their access to the radio channels shared between them and with other pre-existing radio systems by monitoring the channels and reacting responsively to contentions, thereby reducing the overall contentions within the channels.

The distributed deployment of ultra-dense networks is of particular interest in unlicensed spectrum, such as 2.45-GHz and 5.8-GHz industrial, scientific and medical (ISM) radio bands, due to heterogeneous, contention-based wireless network environments with IEEE 802.11 Wi-Fi, IEEE 802.15.1 Bluetooth and IEEE 802.15.4 Zigbee devices prevailing. It is also due to the scarcity of radio communication spectrum. As a matter of fact, the proliferation of Wi-Fi devices, such as access points (APs) and dual-mode smart phones, have already created some form of ultra-dense network in many parts of urban areas, such as hotspots and indoor dense networks.
\begin{figure*}[!t]
\centering
\includegraphics[width=5in]{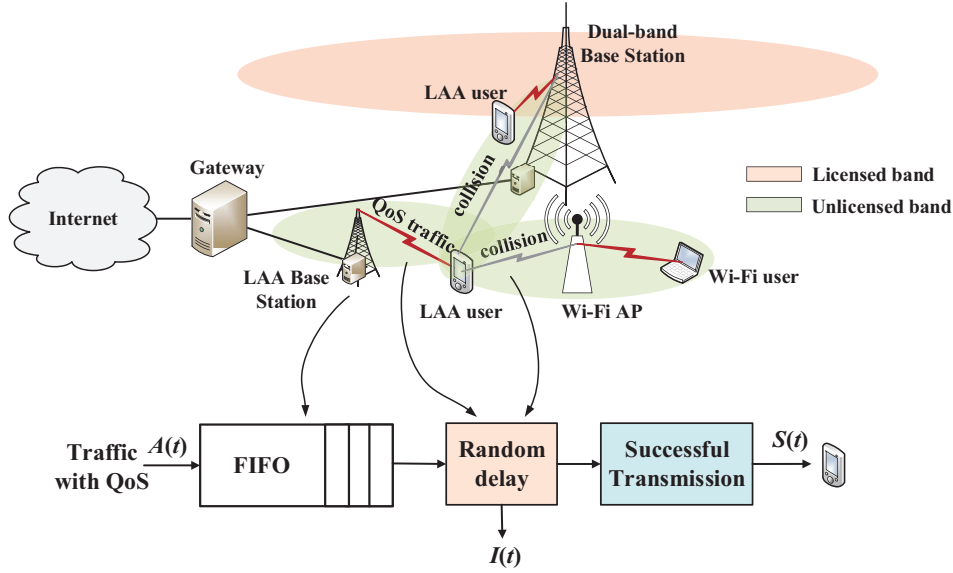}
\caption{The abstraction model, where sophisticated interactions, such as transmission collisions, retransmission backoffs and lossy channel conditions, are all captured in the random delays before successful (re)transmissions.}
\label{fig.tu1}
\end{figure*}

A key issue yet to be addressed for ultra-dense networks
in the contention-based unlicensed spectrum is the trade-off
between capacity and reliability, i.e.\! , quality-of-
service (QoS).
Very little consideration has to date been put on the reliability in contention-based distributed networks. Let alone heterogeneous ultra-dense deployment of such networks~\cite{Cognitivevehicular}. On the other hand, this issue is of paramount importance, as there are extensive discussions to bring in operator-grade networks and services into the unlicensed band, such as the Third Generation Partnership Project (3GPP) LTE and fifth-generation (5G) New Radio (NR). To address the issue, a number of important factors need to be taken into account, as discussed in the following.

\subsection{Heterogeneous Coexistence}
The design goal of ultra-dense networks in the unlicensed band is to comply with regional regulatory requirements, while achieving effective and fair coexistence with legacy systems such as IEEE 802.11 Wi-Fi \cite{3}. Accounting for the coexistence with ubiquitous Wi-Fi networks, listen-before-talk (LBT) techniques are the most promising candidates for ultra-dense networks, where a radio transmitter first senses its radio environment before starting a transmission. In many cases, LBT is implemented in coupling with randomly backing off (re)transmissions, thereby alleviating intrusions towards existing Wi-Fi networks.

Exploiting LBT, 3GPP has specified the license-assisted access (LAA) of long-term evolution (LTE) and 5G systems to the unlicensed band\cite{3}. 3GPP has also considered the use of the Wi-Fi interface to enable a LTE system to access the unlicensed band, as known as the LTE\&Wi-Fi Aggregation (LWA)~\cite{Cui}. As a matter of fact, extensively adopted in Wi-Fi networks, carrier-sense multiple access with collision avoidance (CSMA/CA) is a typical example of LBT, where randomly delayed (re)transmissions are accommodated in exponentially enlarging contention-windows (CWs) adapting to (re)transmission collisions~\cite{wang2015vanet}.

\subsection{Capacity}

To increase network capacity is the primarily objective of the ultra-dense network deployment in the unlicensed band, as mentioned earlier.
Exploiting LBT, two popular contention-based LAA protocols are LBT with fixed CW and LBT with an exponentially increasing CW, referred to as Fixed CW (FCW) and Variable CW (VCW), respectively, as specified in 3GPP TR 36.889 \cite{3}.
These protocols also remain the strongest candidates for the upcoming 5G cellular communication standard on the exploitation of the unlicensed spectrum~\cite{NR}.
Originally designed to offload excessive traffic for the downlinks from the licensed bands, the protocols have also been considered for the uplink applications in the unlicensed band~\cite{3}.
Different from Wi-Fi APs, the underlaying network infrastructure can coordinate between LAA stations in the unlicensed bands and base stations in licensed bands. Signaling, such as channel state information (CSI) and channel quality indictor (CQI), can also be instantly fed back from user terminals to LAA stations, through the licensed band, thereby facilitating fast power control and user scheduling of LAA in the unlicensed bands.

Various published results indicate that VCW can improve the coexistence between LAA and Wi-Fi, from the perspective of capacity~\cite{BChen}. Using VCW and FCW, LAA can outperform LWA in terms of capacity by configuring LBT parameters such as channel sensing time (also known as channel clearance assess, or CCA), CW size, and frame duration, to give priority to LAA over Wi-Fi in the unlicensed band~\cite{3}. In contrast, LWA employs the same air interface as Wi-Fi, and no priority is given to LWA or Wi-Fi~\cite{Cui}.

\subsection{Reliability}
Future wireless networks are expected to support a variety of emerging services, such as high-definition video, three-dimensional visualization, augmented reality, and machine type communications (MTC), where reliability (or in other words, QoS) is crucial.
These types of traffic have typically stringent deadlines to keep traffic flows consistent.
For example, voice over IP (VoIP) can only tolerate packet delays of up to 50ms, while MTC for factory automation requires the delay to be less than 100 $\mu$s and the packer error probability (PER) to be no larger than $10^{-9}$~\cite{NA}.

Unfortunately, it is difficult to preserve reliability in the unlicensed spectrum which is primarily dominated by uncoordinated transmissions of contention-based wireless devices.
Erratic collisions between the uncoordinated transmissions can breach the delay requirements of traffic, and even cause significant packet losses. The ultra-dense deployment of networks could further deteriorate the reliability in the unlicensed spectrum. Moreover, there is typically no policy to regulate the deployment of wireless transmitters in the unlicensed band, given the distributed nature of the networks and devices.

\subsection{Capacity versus Reliability}
As discussed, to leverage between the capacity and reliability is of paramount importance to future networks, but extremely challenging to heterogeneous ultra-dense networks in the unlicensed band, due to the primary capacity boosting objective and the distributed nature of the networks. As a matter of fact, the reliability remains yet to be addressed in Wi-Fi, though Wi-Fi has been around for decades. The latest versions of Wi-Fi, such as IEEE 802.11e enhanced distributed channel access (EDCA), attempting to incorporate QoS, essentially provide relative priorities and cannot guarantee QoS. The QoS of EDCA can deteriorate drastically as the networks get dense and collisions become intensive.

This article advocates a new measure of link capacity which is able to preserve the reliability of the link, thereby accommodating both of these aspects in a single QoS framework. This starts by revisiting the definition of QoS, and identifies the statistic characteristics of QoS suitable for distributed networks. By exploiting the statistical characterization of QoS, a new QoS-preserving capacity is put forth to measure and quantify the maximum consistent data rate of a link without violating the QoS in heterogeneous ultra-dense distributed networks. Potential practical applications of the new measure are also discussed with examples provided.

\section{Reliability-preserving Effective Capacity}

A different way of defining the capacity, known as ``effective capacity'', has been developed to measure the capacity of a link while preserving the reliability (or QoS) of the link~\cite{20}. More specifically, the effective capacity is able to quantify the maximum, consistent transmit rate that a wireless link can sustain given QoS requirements of the link.

Different from existing QoS measures, the QoS requirements of a link, namely, $\{ D_{\max },P_{\rm th}\} $, are parameterized under the concept of effective capacity by a QoS exponent ${\theta}$, which specifies the exponentially decaying rate of the probability $P_{\rm th}$ that the delay threshold $D_{\max }$ is exceeded. A larger $\theta$ corresponds to a more stringent delay requirement.
Denoted by $C(\theta)$, the effective capacity specifies the maximum, consistent, steady-state arrival rate at the input of the First-In-First-Out (FIFO) queue without violating the QoS requirement parameterized by $\theta$, as given by \cite{20,21}
\begin{equation}\label{effective capacity}
C(\theta) = - \mathop {\lim }\limits_{t \to \infty } \frac{1}{{\theta t}}\log (\mathbb{E}\big\{{e^{ - \theta S(t)}}\big\} ),
\end{equation}
and the delay-bound violation probability of the link can be given by ${P_{\rm th}} \approx  {\eta}{e^{ - {\theta}{C}({\theta }){D_{\max }}}}$ \cite{21}, where $S(t)$ denotes the successfully delivered packets during the time period $(0,t]$, ${\eta}$ is the probability of a non-empty queue and can be approximated as the ratio of the constant arrival rate to the average transmit rate. $\mathbb{E}\{\cdot\}$ denotes expectation.


We note that, distinctively different from existing QoS frameworks where resources are scheduled to satisfy QoS requirements, the effective capacity of \eqref{effective capacity} quantifies the maximum QoS-preserving capacity of a radio link. This is interesting in heterogeneous ultra-dense distributed networks, due to the fact that the contention-based, distributed networks are interactive through collisions. Conventional resource scheduling cannot cope with the collisions and the consequent interactions in the absence of centralized coordination in the unlicensed band.

\subsection{Effective Capacity of Point-to-Point Link}
Until very recently, the idea of effective capacity has been adopted in single point-to-point link~\cite{20}, or homogeneous network environments such as cellular systems~\cite{21}.
Even though these scenarios are substantially different from the heterogeneous ultra-dense networks of interest, some interesting properties of effective capacity can still be revealed. For instance, the effective capacity is a monotonically decreasing function of $\theta$, which implies that more stringent QoS requirements would result in lower supportable service rates. As $\theta$ approaches zero, the service does not impose any constraint on the queue length and delay bound, and thus the effective capacity converges to the conventional capacity without QoS consideration. In contrast, when ${\theta }$ approaches infinity, i.e., implying the zero delay requirement, the effective capacity degrades to the minimum service rate over all channel fading states.

\subsection{Heterogeneous Ultra-dense Distributed Networks}
It is non-trivial to measure the effective capacity of heterogeneous, ultra-dense distributed networks in the presence of transmission collisions and lossy wireless channels though.
One reason is because the transmission collisions can result in non-deterministic and unpredictable delays. Moreover, the delays are also time-variant. Particularly, a transmitter can delay its (re)transmission by a random number of time slots to alleviate collisions, while other nodes' transmissions can prolong some of the slots and further randomize the delay before the designated transmitter starts to send.
For these reasons, typical techniques of analyzing the capacity of distributed networks, such as Markov model, cannot apply.

\begin{figure}[!t]
\centering
\includegraphics[width=3.8in]{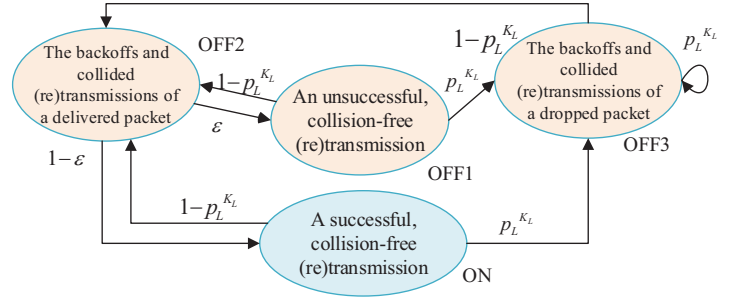}
\caption{A new four-state semi-Markovian model, which is able to quantify the effective capacity of the link specified in Fig. 1(a) in a heterogeneous distributed network environment by modeling the random delays.}
\label{fig.3}
\end{figure}
Our previous study \cite{Cui2017effective}, has attempted to interpret the wireless channel between a transmitter-receiver pair in a heterogeneous contention-based network with $N$ LAA transmitters and $M$ Wi-Fi transmitters as a FIFO queue with random perturbation/delay and loss, as illustrated in Fig. \ref{fig.tu1}, where $A(t)$ denotes the packet arrival and $I(t)$ collects the packets that are dropped after their maximum number of retransmissions exceeded, both during the time period $(0,t]$. $S(t)$ denotes the successfully delivered packets during the time period, as defined earlier.

With this interpretation, a new semi-Markovian ON/OFF model can be developed to capture the random perturbation/delay and loss at the output of the FIFO queue, as shown in Fig. \ref{fig.3}. There are four states as an uncoordinated transmitter can experience. An ON state accounts for a collision-free and successful packet (re)transmission. An OFF$_1$ state accounts for a collision-free yet unsuccessful (re)transmission due to lossy wireless channels. An OFF$_2$ state captures the variable delay prior to a collision-free (re)transmission, and an OFF$_3$ state captures the variable delay caused by a packet that exhausts all retransmissions with collisions before getting dropped.

Different from classic Markov models with deterministic states, a semi-Markovian model consists of states with random variable durations, as the OFF$_2$ and OFF$_3$ states in the four-state model depicted in Fig. \ref{fig.3} and can be generally described by the state transition matrix $\mathbf{P}$ and the moment generating functions (MGFs) of the variable durations of the states.
The effective capacity of a semi-Markovian model, i.e., $C(\theta)$, as defined in \eqref{effective capacity}, has been proved to be the unique solution for $\rho(\theta,C(\theta))=1$, where $\rho(\theta,C(\theta))$ is the spectral radius (i.e., the largest eigenvalue) of $\mathbf{\Gamma}(\theta,C(\theta))\mathbf{P}$, and $\mathbf{\Gamma}(\theta,C(\theta))$ is the diagonal matrix collecting the MGFs of the variable durations. This has been applied to and validated in, a two-state ON/OFF fluid source which emits and withholds emissions in the ON and OFF states with variable durations, respectively~\cite{kontovasilis1997effective}.

In the context of heterogeneous ultra-dense networks, the MGFs of the delays of the OFF$_2$ and OFF$_3$ states can be evaluated by using classic Markovian techniques. The transition matrix $\mathbf{P}$ can be straightforwardly obtained, as shown in Fig. \ref{fig.3}.
By evaluating $\rho(\theta,C(\theta))=1$, an important conclusion has been drawn \cite{Cui2017effective}. Specifically, in the presence of $N$ LAA transmitters and $M$ Wi-Fi transmitters, the effective capacity of a LAA transmitter, $C(\theta)$, can be given by~\cite[Theorem 1]{Cui2017effective}
\begin{equation}\label{eq29}
\begin{aligned}
(1 - p_L^{{K_L}})&{{\hat t}_1}({e^{{\theta}{C}}})\Big[{e^{( - {R}{\theta} + {\theta }{C}){T_{\rm f}}}}(1 - {\varepsilon})\\
& + {e^{{\theta}{C}{T_{\rm f}}}}{\varepsilon}\Big] + p_L^{{K_L}}{{\hat t}_2}({e^{{\theta }{C}}}) = 1,
\end{aligned}
\end{equation}
where $\hat{t}_{1}(\cdot)$ and $\hat{t}_{2}(\cdot)$ are the probability generation functions (PGFs) of the durations of backoffs for a delivered packet and those for a dropped packet, respectively; $p_L$ is the collision probability of the designated LAA transmitter; $T_{\rm f}$ is the duration of a collision-free (re)transmission of the transmitters; $\varepsilon$ is the PER of collision-free (re)transmissions of the transmitter; and $R$ is the instantaneous transmit rate of the transmitter.

\section{Density and Insight of Heterogeneous Distributed Networks}
\begin{figure}[t]
\centering
\subfigure[]{
\includegraphics[width=3.5in]{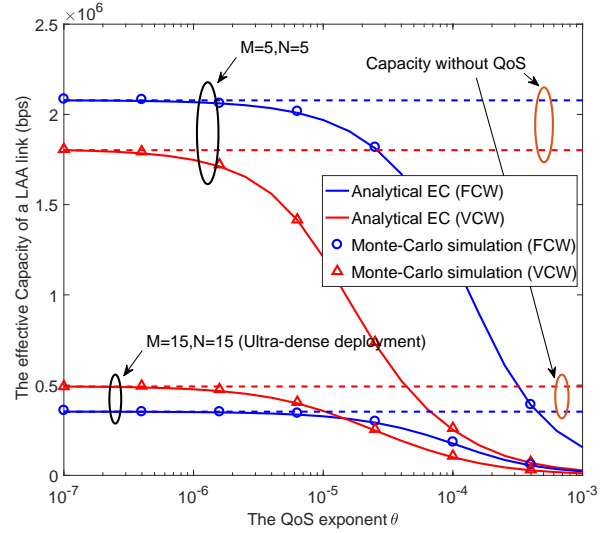}
\label{figg1}
}
\subfigure[]{
\includegraphics[width=3.5in]{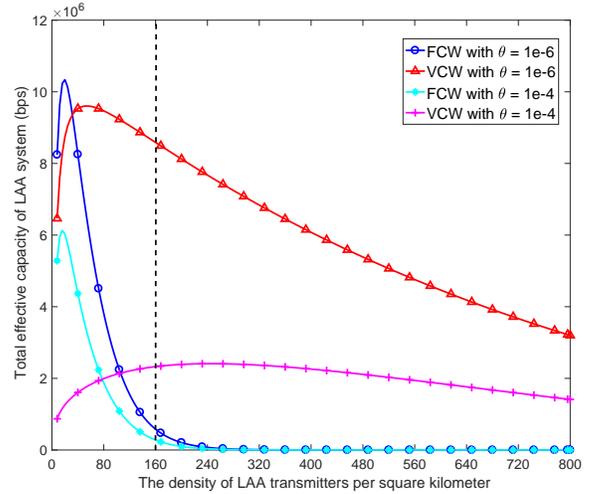}
\label{figg41}
}
\vspace{-0.2cm}
\caption{(a) The effective capacity of a link vs $\theta$, (b) Total effective capacity vs the density of LAA transmitters, given the number of Wi-Fi $M=5$.}
\end{figure}

The new results of considering the effective capacity has been validated by comparing with the well-studied, conventional capacity with no QoS consideration.
Simulations are carried out in a network with $N$ LAA transmitters and $M$ Wi-Fi transmitters. The transmit powers of the LAA and Wi-Fi transmitters are 23 dBm. The system bandwidth is 5MHz.
As shown in Fig. \ref{figg1}, the QoS-preserving effective capacity indistinguishably converges to the capacity in both cases of FCW and VCW, as $\theta$ approaches to zero, i.e., the link has no QoS requirement. In other words, the effective capacity recedes to the capacity in the case where the QoS requirement is too loose. On the other hand, when $\theta$ is large, the effective capacity asymptotically approaches to zero. This is due to the fact that the QoS requirement becomes too tight to support any meaningful consistent data rate.

Unexpected observations have been made under the reliability consideration, distinctively different from various earlier results with no such consideration \cite{BChen}.
Particularly, Fig. \ref{figg1} shows that FCW can significantly outperform VCW in the presence of a small number of Wi-Fi nodes and stringent QoS requirements, though VCW has been shown to outperform FCW in terms of coexistence ~\cite{BChen}. One reason is because VCW can suffer from severe exponentially delayed retransmissions, violate QoS requirements, and therefore incur significant loss of effective capacity. Another reason is because FCW, sticking to a fixed small size of CW, gives priority to LAA transmitters and becomes intrusive to Wi-Fi, as compared to VCW. However, FCW is less effective in terms of reacting to intensive collisions, as also shown in the Fig. \ref{figg1}, and therefore is bypassed by VCW in the presence of large numbers of active transmitters in its neighborhood.

Capturing fine-grained QoS requirements of future 5G, the new effective capacity theory can play a key role of identifying appropriate access techniques and predicting adequate network densities of 5G in the heterogeneous distributed network environment of the unlicensed band. The selection of FCW or VCW is crucial to the QoS-preserving effective capacity as shown in Fig. \ref{figg41}, and is susceptible to the density of the heterogeneous network in the unlicensed band. We can see that the total effective capacity of LAA system increases with its the density, when the density is relatively low. This is due to the fact that the interference or contention is limited and increasing the network density helps efficiently utilize the channel. We also see that the effective capacity decreases as the network density keeps growing, since the collisions of uncoordinated transmissions become intensive between heterogeneous devices

Fig. \ref{figg41} also shows the effective capacity of VCW decreases far more slowly than that of FCW. Furthermore, when the density is 32 transmitters per square kilometer, the effective capacity of VCW surpasses that of FCW in the case of $\theta=10^{-6}$ and the network density of 32 transmitters per square kilometer. The reason is that VCW is superior to FCW in terms of the coexistence not only with Wi-Fi devices, but also between LAA transmitters, especially in dense network environments. To this end, VCW can provide far more effective support for ultra-dense network deployments in heterogeneous distributed network environments, such as those in the unlicensed band. Take $\theta=10^{-6}$ and the density as 160 transmitters per square kilometer for an example. VCW is able to support more than eighteen times the capacity of FCW.

\section{Practical Applications of Effective Capacity}\label{sec: practical value}

Apart from its analytical value, the effective capacity, developed for heterogeneous ultra-dense distributed networks, is also versatile and, like the conventional (Shannon) capacity, can be readily used to configure and optimize network operations. In particular, despite the sophisticated expression given in \eqref{eq29} (as opposed to a simple closed form like the capacity), the effective capacity $C(\theta)$ can be rigorously proved to be strictly concave with respect to the instantaneous transmit rate $R$~\cite[Theorem 2]{Cui2017effective}.
The effective capacity can be used to facilitate power control, resource allocation and admission control (as the Shannon capacity) of traffic flows with non-trivial QoS requirements, yet unprecedentedly in heterogeneous ultra-dense distributed network environments in the unlicensed spectrum.

\subsection{Effective Capacity Region and Admission Control}
\begin{figure}[t]
\centering
\includegraphics[width=3.5in]{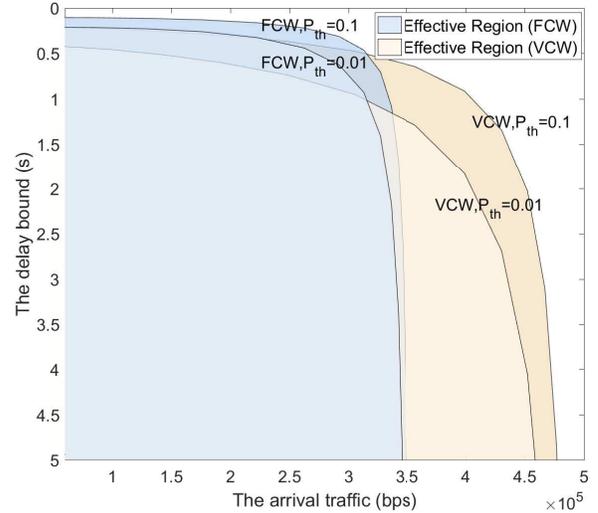}
\caption{Delay bound vs the traffic, where $M=15$ and $N=15$.}
\label{fig.21}
\end{figure}

Using the new result of \eqref{eq29}, the important region of the effective capacity can be plotted, as done in Fig. \ref{fig.21}, and the trade-off between the effective capacity and the delay bound $D_{\max}$ can be measured. Given the network density and the delay-bound violation probability threshold $P_{\rm th}$, the reliability and effective capacity that a transmitter can achieve in a heterogeneous distributed network are captured in the corresponding shaded area enclosed by a curve plotted by using \eqref{eq29}, as well as the two axises. The shaded area, or region, enlarges, as $P_{\rm th}$ increases (i.e., the QoS requirement gets loose). Given the network density and $P_{\rm th}$, VCW can generally have a larger region than FCW. Nevertheless, FCW can typically better support stringent QoS in the low range of traffic arrival rate, while VCW is more suitable for traffic with relatively loose QoS requirements.

This region of the effective capacity versus delay bound sheds important insights to understand and predict the capability of a heterogeneous ultra-dense distributed network in terms of capacity and reliability.
The region is also of practical value to make adequate admission control decisions for new traffic arrivals with non-trivial QoS requirements. Important parameters, such as network density, channel condition, QoS requirement and minimum data rate, can all be captured while such a decision is being made under the guidance of the region.

\subsection{QoS-aware Power Control}
The maximization of the effective capacity in a heterogeneous ultra-dense distributed network can also be formulated by optimizing the power of a transmitter concurrently sending data to multiple receivers using orthogonal frequency-division multiple-access (OFDMA) techniques. By exploiting the aforementioned concavity of the effective capacity, the problem can be reformulated to be a convex optimization problem which can be straightforwardly solved by using standard convex techniques, such as interior-point method and sub-gradient method~\cite{Cui2017effective}.

\begin{figure}[t]
\centering
\includegraphics[width=3.5in]{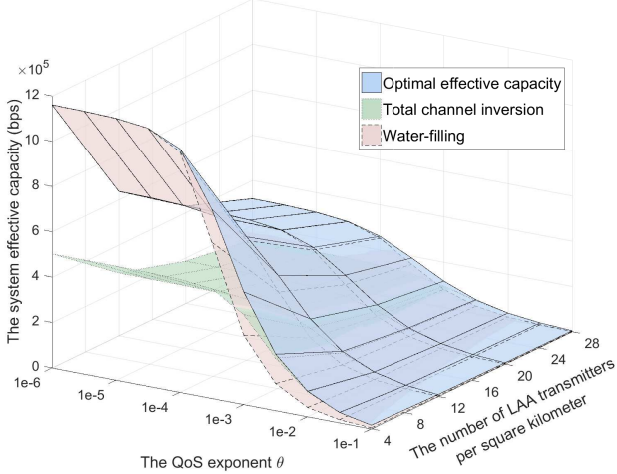}
\caption{The effective capacity for different power allocation strategies, where there are $M=4$ Wi-Fi transmitters, and $20$ LAA receivers per transmitter, and all LAA receivers are assumed to have the same delay and PER requirements, i.e., $\theta_1=\cdots=\theta_K$, for illustration convenience.}
\label{fig611}
\end{figure}

Fig. \ref{fig611} corroborates the effectiveness of the use of the effective capacity for power control, where there are 20 LAA receivers per LAA transmitter, and $M = 4$ Wi-Fi transmitter; the system bandwidth is $B=20$ MHz; and VCW is adopted. For comparison purpose, a water-filling method \cite{Goldsmith1997} and a total channel inversion method \cite{Tang2007} are also simulated. As the state of the art, the water-filling method only maximizes the capacity by using the channel gains. The total channel inversion method allocates the transmit power inversely proportionally to the channel gain of every LAA receiver. The method is asymptotically optimal for maximizing the effective capacity of a single wireless point-to-point link as $\theta$ approaches infinity, as proved in \cite{Tang2007}.

Fig. \ref{fig611} shows that the new technique, that maximizes the effective capacity based on the new result \eqref{eq29}, is able to increasingly outperform the water-filling method, as $\theta$ increases (i.e., the QoS becomes increasingly stringent). For instance, the gain of the technique is up to 62.7\% in the case of $\theta=10^{-3}$ and the density as 4 per square kilometer. On the other hand, the new technique is indistinguishably close to water-filling, when $\theta\rightarrow 0$. This is because the effective capacity recedes to the capacity under loose QoS, while water-filling maximizes the capacity. Further, the new effective capacity based technique, is able to outperform the total channel inversion method across a wide spectrum of $\theta\leq 10^{-2}$. For $\theta>10^{-2}$, the new technique provides the same performance as the total channel conversion method which is asymptotically optimal as $\theta\rightarrow \infty$.

\subsection{QoS-aware Bandwidth Allocation}
The effective capacity can also be used to facilitate the
bandwidth allocation in the case that a LAA transmitter serves multiple receivers with non-trivial QoS requirements, as discussed in Sections IV-B. Exploiting the aforementioned concavity of the effective capacity with respect to the instantaneous transmit rate, the bandwidth allocation can be formulated as a convex optimization problem which is readily solved by using standard convex techniques.

\begin{figure}[t]
\centering
\includegraphics[width=3.2in]{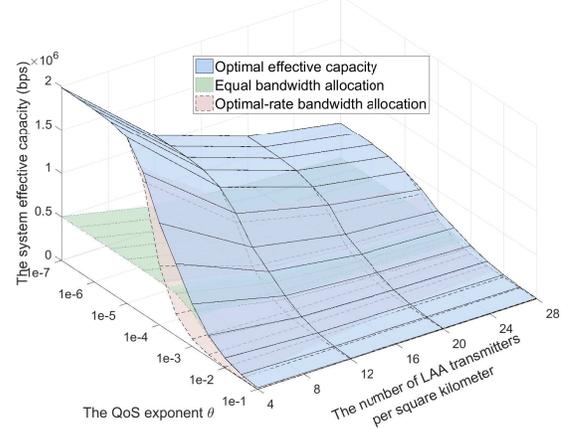}
\vspace{-0.4cm}
\caption{The effective capacity for different bandwidth allocation strategies, where there are $M=4$ Wi-Fi transmitters, and $20$ LAA receivers per transmitter, and all LAA receivers are assumed to have the same delay and PER requirements, i.e., $\theta_1=\cdots=\theta_K$, for illustration convenience.}
\label{fig61}
\vspace{-0.2cm}
\end{figure}

Fig. \ref{fig61} shows that the bandwidth allocation based on the effective capacity is superior to other existing strategies, namely, ``optimal-rate bandwidth allocation'' and ``equal bandwidth allocation'', where there are 20 LAA receivers per transmitter, and $M = 4$ Wi-Fi transmitter; the system total bandwidth is $B=20$ MHz; and VCW is adopted. The optimal-rate bandwidth allocation performs the bandwidth allocation to maximize the capacity with no QoS consideration. The equal bandwidth allocation is self-explanatory. We also assume all LAA receivers have the same QoS requirements, and independent and identically distributed channels.

In the case that $\theta\rightarrow 0$, the effective capacity recedes to the capacity, as discussed. The new effective capacity based technique is able to maximize the capacity as the optimal-rate bandwidth allocation can, and significantly outperform the equal bandwidth allocation. In the case that $\theta\rightarrow \infty$, the new technique is able to substantially bypass the optimal-rate method. The new technique is indistinguishably close to the equal resource allocation, since the equal resource allocation is known to provide consistent support for non-trivial QoS as the cost of capacity. In this sense, the effective capacity is more sensitive to the QoS stringency than it is to the link quality. Our design is of great practical value under a wide range of medium to stringent QoS conditions.

\subsection{Other Applications}
Another application example of the effective capacity is to maximize the effective energy efficiency of transmitters in heterogeneous ultra-dense distributed networks. This is an extension of the above effective capacity maximization, since the effective energy efficiency of a transmitter is defined to be the ratio of its effective capacity to its transmit power. By exploiting the aforementioned concavity of the effective capacity again, as well as fractional programming techniques, the maximization of the effective energy efficiency can be formulated to be a parametric convex problem and recursively solved by using the KKT conditions and the Dinkelbach's method~\cite{Cui2017effective}.

\section{Conclusion}
This article discusses the key dilemma between capacity and reliability in heterogeneous, ultra-dense, distributed networks residing in the unlicensed spectrum. The new measure of effective capacity is advocated, which quantifies the maximum consistent data rate of a link without violating the reliability (or QoS) of the link.
Particularly, we present the recent non-trivial development of the effective capacity in heterogeneous ultra-dense distributed networks, as well as its potential applications to the admission control, power control and resource allocation of practical networks. Examples of the applications are provided, with substantial gains revealed as compared to existing technologies.
\section*{Acknowledgment}
The work was supported by National Nature Science Foundation of China Project (Grant No. 61471058), International Cooperation NSFC Program (61461136002), the National High Technology Program of China (2014AA01A701), Hong Kong, Macao and Taiwan Science and Technology Cooperation Projects (2016YFE0122900) and the 111 Project of China (B16006).

\bibliographystyle{IEEEtran}

\begin{IEEEbiographynophoto}{Qimei Cui}
(M'09-SM'15) (cuiqimei@bupt.edu.cn)  received her Ph.D. degree from Beijing University of Posts and Telecommunications (BUPT), Beijing, China in 2006. She was a full professor in School of Information and Communication Engineering at BUPT since 2014. She was guest professor in department of Electronic Engineering of University of Notre Dame in 2016. Her main research interests include spectral-efficiency or energy-efficiency based transmission theory and networking technology for 4G/5G networks and green communications.
\end{IEEEbiographynophoto}
\vspace{-1cm}
\begin{IEEEbiographynophoto}{Yu Gu} received the B.S. degree in communication engineering from Dalian University of Technology, Dalian, China, in 2014. He is currently working toward the Ph.D. degree in communications and information systems with the Beijing University of Posts and Telecommunications (BUPT), Beijing, China. His main research interests include stochastic optimization theory, machine learning, large-scale distributed systems as well as their applications to 5G heterogeneous wireless networks.
\end{IEEEbiographynophoto}
\vspace{-1cm}
\begin{IEEEbiographynophoto}{Wei Ni} (M'09-SM'15) received the B.E. and Ph.D. degrees in Electronic Engineering from Fudan University, Shanghai, China, in 2000 and 2005, respectively. Currently he is a Senior Scientist, Team and Project Leaders at CSIRO, Australia. He also holds adjunct positions at the University of New South Wales (UNSW), Macquarie University (MQ) and the University of Technology Sydney (UTS). Prior to this he was Deputy Project Manager at the Bell Labs R\&I Center, Alcatel/Alcatel-Lucent (2005-2008).
\end{IEEEbiographynophoto}
\vspace{-1cm}
\begin{IEEEbiographynophoto}{Xuefei Zhang} received the B.S. and Ph.D. degrees in telecommunications engineering from the Beijing University of Posts and Telecommunications (BUPT) in 2010 and 2015, respectively. She is currently with the National Engineering Lab, BUPT. Her research area includes heterogeneous network, machine technology communications, stochastic geometry, and optimization theory.
\end{IEEEbiographynophoto}
\vspace{-1cm}
\begin{IEEEbiographynophoto}{Xiaofeng Tao} [M'05-SM'13] (taoxf@bupt.edu.cn) received his B.S degree from Xi'an Jiaotong University, China, in 1993, and M.S. and Ph.D. degrees from BUPT in 1999 and 2002, respectively. He is an IET Fellow. He was chief architect of the Chinese National Future 4G TDD working group from 2003 to 2006 and established the 4G TDD CoMP trial network in 2006. He is currently a professor at BUPT.
\end{IEEEbiographynophoto}
\vspace{-1cm}
\begin{IEEEbiographynophoto}{Ping Zhang} (pzhang@bupt.edu.cn) received his Ph.D. degree from BUPT in 1990. Now he is a professor at BUPT and vice director of the Ubiquitous Networking Task Commission of the Chinese Communication Standardization Association. He is the chief scientist of the National Program on Key Basic Research Project (973 Program) and also serves as a member of the China 3G Group China 863 FuTURE project.
\end{IEEEbiographynophoto}
\vspace{-1cm}
\begin{IEEEbiographynophoto}{Ren Ping Liu} (M'09-SM'14)is a Professor at the School of Computing and Communications in University of Technology Sydney. Prior to that he was a Principal Scientist at CSIRO, where he led wireless networking research activities. He specialises in protocol design and modelling, and has delivered networking solutions to a number of government agencies and industry customers. Professor Liu was the winner of Australian Engineering Innovation Award and CSIRO Chairman's medal.
\end{IEEEbiographynophoto}

\end{document}